\documentclass[apjl,ip,twocolumn,trackchanges]{aastex631}
\usepackage{comment}
\usepackage{ifthen}
\usepackage[switch]{lineno}
\modulolinenumbers[5]

\newcommand{\forloop}[5][1]%
{%
\setcounter{#2}{#3}%
\ifthenelse{#4}%
	{%
	#5%
	\addtocounter{#2}{#1}%
	\forloop[#1]{#2}{\value{#2}}{#4}{#5}%
	}%
	{%
	}%
}%

\newcommand{\ctbd}[1]{}

\newcommand{\Lc}{Light curve}

\newcommand{\ccdsize}[1]{\ensuremath{\rm #1\times\rm#1}}

\newcommand{\tsize}[1]{\mbox{\rm #1 m}}

\newcommand{\masy}{\ensuremath{\rm mas\,yr^{-1}}}
\newcommand{\kms}{\ensuremath{\rm km\,s^{-1}}}
\newcommand{\ms}{\ensuremath{\rm m\,s^{-1}}}

\newcommand{\gcmc}{\ensuremath{\rm g\,cm^{-3}}}

\newcommand{\C}{\ensuremath{^{\circ}C\;}}

\newcommand{\teff}{\ensuremath{T_{\rm eff}}}

\newcommand{\feh}{\ensuremath{\rm [Fe/H]}}

\newcommand{\rsun}{\ensuremath{R_\sun}}
\newcommand{\msun}{\ensuremath{M_\sun}}
\newcommand{\lsun}{\ensuremath{L_\sun}}

\newcommand{\rstar}{\ensuremath{R_\star}}
\newcommand{\mstar}{\ensuremath{M_\star}}
\newcommand{\lstar}{\ensuremath{L_\star}}

\newcommand{\teffstar}{\ensuremath{T_{\rm eff\star}}}
\newcommand{\rhostar}{\ensuremath{\rho_\star}}
\newcommand{\loggstar}{\ensuremath{\log{g_{\star}}}}

\newcommand{\rpl}{\ensuremath{R_{p}}}
\newcommand{\mpl}{\ensuremath{M_{p}}}

\newcommand{\rhopl}{\ensuremath{\rho_{p}}}

\newcommand{\arstar}{\ensuremath{a/\rstar}}
\newcommand{\zrstar}{\ensuremath{\zeta/\rstar}}

\newcommand{\rjup}{\ensuremath{R_{\rm J}}}
\newcommand{\mjup}{\ensuremath{M_{\rm J}}}

\newcommand{\refsecl}[1]{\mbox{Section \ref{sec:#1}}}

\newcommand{\hatcurhtr}{TOI-3235}                          
\newcommand{\hatcurCCra}{\ensuremath{13^{\mathrm h}49^{\mathrm m}53.9777{\mathrm s}}}                   
\newcommand{\hatcurCCdec}{\ensuremath{-46{\arcdeg}03{\arcmin}58.4541{\arcsec}}}                 
\newcommand{\hatcurCCtwomass}{2MASS~13495398-4603583}     
\newcommand{\hatcurCCgaiadrthree}{GAIA~DR3~6107144260251920000} 
\newcommand{\hatcurTICID}{TIC~243641947}                  
\newcommand{\hatcurCCparallax}{\ensuremath{13.781\pm0.027}} 
\newcommand{\hatcurCCgaiamGthree}{\ensuremath{14.4605\pm0.0028}} 
\newcommand{\hatcurCCgaiamBPthree}{\ensuremath{15.9000\pm0.0042}} 
\newcommand{\hatcurCCgaiamRPthree}{\ensuremath{13.2881\pm0.0038}} 
\newcommand{\hatcurCCtwomassJmag}{\ensuremath{11.706\pm0.025}} 
\newcommand{\hatcurCCtwomassHmag}{\ensuremath{11.099\pm0.024}} 
\newcommand{\hatcurCCtwomassKmag}{\ensuremath{10.819\pm0.021}} 
\newcommand{\hatcurCCWonemag}{\ensuremath{10.694\pm0.023}} 
\newcommand{\hatcurCCWtwomag}{\ensuremath{10.548\pm0.021}} 
\newcommand{\hatcurCCWthreemag}{\ensuremath{10.458\pm0.064}} 
\newcommand{\hatcurLCrprstar}{\ensuremath{0.2828\pm0.0016}} 
\newcommand{\hatcurLCbsq}{\ensuremath{0.261_{-0.012}^{+0.012}}} 
\newcommand{\hatcurLCimp}{\ensuremath{0.511_{-0.012}^{+0.011}}} 
\newcommand{\hatcurLCzeta}{\ensuremath{44.41\pm0.32}}     
\newcommand{\hatcurLCdur}{\ensuremath{0.06165\pm0.00021}} 
\newcommand{\hatcurLCingdur}{\ensuremath{0.01765\pm0.00030}} 
\newcommand{\hatcurLCP}{\ensuremath{2.59261842\pm0.00000041}} 
\newcommand{\hatcurLCPshort}{\ensuremath{2.5926}}         
\newcommand{\hatcurLCT}{\ensuremath{2459690.001730\pm0.000045}} 
\newcommand{\hatcurLCiblendA}{\ensuremath{0.982\pm0.022}} 
\newcommand{\hatcurLCiblendB}{\ensuremath{0.957\pm0.027}} 
\newcommand{\hatcurLCiblendC}{\ensuremath{1.084\pm0.011}} 
\newcommand{\hatcurLCiblendD}{\ensuremath{1.1248\pm0.0088}} 
\newcommand{\hatcurLCrho}{\ensuremath{10.99\pm0.13}}      
\newcommand{\hatcurSMEiteff}{\ensuremath{3196\pm67}}      
\newcommand{\hatcurSMEizfeh}{\ensuremath{-0.02\pm0.10}}   
\newcommand{\hatcurLBig}{\ensuremath{0.31\pm0.12}}        
\newcommand{\hatcurLBiig}{\ensuremath{0.35\pm0.14}}       
\newcommand{\hatcurLBir}{\ensuremath{0.41\pm0.15}}        
\newcommand{\hatcurLBiir}{\ensuremath{0.32\pm0.16}}       
\newcommand{\hatcurLBii}{\ensuremath{0.30\pm0.11}}        
\newcommand{\hatcurLBiii}{\ensuremath{0.26_{-0.15}^{+0.11}}} 
\newcommand{\hatcurLBiz}{\ensuremath{0.140_{-0.081}^{+0.106}}} 
\newcommand{\hatcurLBiiz}{\ensuremath{0.17\pm0.13}}       
\newcommand{\hatcurLBiT}{\ensuremath{0.24\pm0.12}}        
\newcommand{\hatcurLBiiT}{\ensuremath{0.33\pm0.16}}       
\newcommand{\hatcurISOm}{\ensuremath{0.3939\pm0.0030}}    
\newcommand{\hatcurISOmlong}{\ensuremath{0.3939\pm0.0030}} 
\newcommand{\hatcurISOrlong}{\ensuremath{0.3697\pm0.0018}} 
\newcommand{\hatcurISOlogg}{\ensuremath{4.8976\pm0.0035}} 
\newcommand{\hatcurISOlum}{\ensuremath{0.01623\pm0.00018}} 
\newcommand{\hatcurISOteff}{\ensuremath{3388.8\pm5.9}}    
\newcommand{\hatcurISOteffshort}{\ensuremath{3389}}    
\newcommand{\hatcurISOzfeh}{\ensuremath{0.264_{-0.017}^{+0.013}}} 
\newcommand{\hatcurISOage}{\ensuremath{0.394_{-0.090}^{+0.152}}} 
\newcommand{\hatcurRVK}{\ensuremath{182.9\pm3.3}}         
\newcommand{\hatcurRVjittertwosiglim}{\ensuremath{<9.6}}  
\newcommand{\hatcurPPi}{\ensuremath{88.140\pm0.046}}      
\newcommand{\hatcurPPlogg}{\ensuremath{3.202\pm0.041}}    
\newcommand{\hatcurPPar}{\ensuremath{15.75\pm0.73}}     
\newcommand{\hatcurPParel}{\ensuremath{0.02709\pm0.00046}} 
\newcommand{\hatcurPPrho}{\ensuremath{0.78\pm0.11}}     
\newcommand{\hatcurPPm}{\ensuremath{0.665\pm0.025}}       
\newcommand{\hatcurPPmlong}{\ensuremath{0.665\pm0.025}}   
\newcommand{\hatcurPPr}{\ensuremath{1.017\pm0.044}}     
\newcommand{\hatcurPPrlong}{\ensuremath{1.017\pm0.044}} 
\newcommand{\hatcurPPmrcorr}{\ensuremath{0.09}}           
\newcommand{\hatcurPPteff}{\ensuremath{604\pm19}}      
\newcommand{\hatcurPPteffshort}{\ensuremath{604}}      
\newcommand{\hatcurPPtheta}{\ensuremath{0.0896\pm0.0042}} 
\newcommand{\hatcurPPfluxavglog}{\ensuremath{7.479\pm0.018}} 
\newcommand{\hatcurXAv}{\ensuremath{0.064\pm0.021}}       
\newcommand{\hatcurXdistred}{\ensuremath{72.50\pm0.12}}   
\newcommand{\hatcurCCpmra}{\ensuremath{-170.503\pm0.028}} 
\newcommand{\hatcurCCpmdec}{\ensuremath{-64.264\pm0.023}} 
\newcommand{\hatcurRVeccentwosiglimeccen}{\ensuremath{<0.029}} 

\shortauthors{Hobson et al.}
\shorttitle{TOI-3235\lowercase{b}}

\begin{document}

\title{TOI-3235 b: a transiting giant planet around an M4 dwarf star}

\correspondingauthor{Melissa J. Hobson}
\email{hobson@mpia.de}

\author[0000-0002-5945-7975]{Melissa J. Hobson}
\affiliation{Max Planck Institute for Astronomy, K{\"{o}}nigstuhl 17, 69117 - Heidelberg, Germany}
\affiliation{Millennium Institute of Astrophysics (MAS), Nuncio Monseñor Sótero Sanz 100, Providencia, Santiago, Chile}

\author[0000-0002-5389-3944]{Andr\'es Jord\'an}
\affiliation{Facultad de Ingenier\'ia y Ciencias, Universidad Adolfo Ib\'a\~nez, Av.\ Diagonal las Torres 2640, Pe\~nalol\'en, Santiago, Chile}
\affiliation{Millennium Institute of Astrophysics (MAS), Nuncio Monseñor Sótero Sanz 100, Providencia, Santiago, Chile}
\affiliation{Data Observatory Foundation, Santiago, Chile}

\author[0000-0001-7904-4441]{E.~M.\ Bryant}
\affiliation{Department of Physics, University of Warwick, Gibbet Hill Road, Coventry CV4 7AL, UK}
\affiliation{Centre for Exoplanets and Habitability, University of Warwick, Gibbet Hill Road, Coventry CV4 7AL, UK}
\affiliation{Mullard Space Science Laboratory, University College London, Holmbury St Mary, Dorking, Surrey, RH5 6NT, UK}

\author[0000-0002-9158-7315]{R.\ Brahm}
\affiliation{Facultad de Ingenier\'ia y Ciencias, Universidad Adolfo Ib\'a\~nez, Av.\ Diagonal las Torres 2640, Pe\~nalol\'en, Santiago, Chile}
\affiliation{Millennium Institute of Astrophysics (MAS), Nuncio Monseñor Sótero Sanz 100, Providencia, Santiago, Chile}
\affiliation{Data Observatory Foundation, Santiago, Chile}

\author[0000-0001-6023-1335]{D.\ Bayliss}
\affil{Department of Physics, University of Warwick, Gibbet Hill Road, Coventry CV4 7AL, UK}

\author[0000-0001-8732-6166]{J.~D.\ Hartman}
\affil{Department of Astrophysical Sciences, Princeton University, NJ 08544, USA}

\author[0000-0001-7204-6727]{G.\ \'A. Bakos}
\affil{Department of Astrophysical Sciences, Princeton University, NJ 08544, USA}

\author{Th.\ Henning}
\affil{Max Planck Institute for Astronomy, K{\"{o}}nigstuhl 17, 69117 - Heidelberg, Germany}

\author[0000-0003-3208-9815]{Jose Manuel Almenara} 
\affiliation{Univ. Grenoble Alpes, CNRS, IPAG, F-38000 Grenoble, France} 

\author[0000-0003-1464-9276]{Khalid Barkaoui}
\affiliation{Astrobiology Research Unit, Universit\'e de Li\`ege, 19C All\'ee du 6 Ao\^ut, 4000 Li\`ege, Belgium}
\affiliation{Department of Earth, Atmospheric and Planetary Science, Massachusetts Institute of Technology, 77 Massachusetts Avenue, Cambridge, MA 02139, USA}
\affiliation{Instituto de Astrof\'isica de Canarias (IAC), Calle V\'ia L\'actea s/n, 38200, La Laguna, Tenerife, Spain}

\author[0000-0001-6285-9847]{Zouhair Benkhaldoun} 
\affiliation{Oukaimeden Observatory, High Energy Physics and Astrophysics Laboratory, Faculty of sciences Semlalia, Cadi Ayyad University, Marrakech, Morocco \label{ouka}}

\author[0000-0001-9003-8894]{Xavier Bonfils} 
\affiliation{Univ. Grenoble Alpes, CNRS, IPAG, F-38000 Grenoble, France} 

\author{François Bouchy} 
\affiliation{Observatoire de Genève, Département d'Astronomie, Université de Genève, Chemin Pegasi 51b, 1290 Versoix, Switzerland}

\author{David~Charbonneau}
\affiliation{Center for Astrophysics \textbar\ Harvard \& Smithsonian, 60 Garden
  St., Cambridge, MA 02138, USA}

\author{Marion Cointepas}
\affiliation{Univ. Grenoble Alpes, CNRS, IPAG, F-38000 Grenoble, France} 
\affiliation{Observatoire de Genève, Département d'Astronomie, Université de Genève, Chemin Pegasi 51b, 1290 Versoix, Switzerland}

\author[0000-0001-6588-9574]{Karen A.\ Collins}  
\affiliation{Center for Astrophysics \textbar\ Harvard \& Smithsonian, 60 Garden
  St., Cambridge, MA 02138, USA}


\author[0000-0003-3773-5142]{Jason~D.~Eastman} 
\affiliation{Center for Astrophysics \textbar\ Harvard \& Smithsonian, 60 Garden
  St., Cambridge, MA 02138, USA}

\author{Mourad~Ghachoui} 
\affiliation{Astrobiology Research Unit, Universit\'e de Li\`ege, 19C All\'ee du 6 Ao\^ut, 4000 Li\`ege, Belgium}
\affiliation{Oukaimeden Observatory, High Energy Physics and Astrophysics Laboratory, Faculty of sciences Semlalia, Cadi Ayyad University, Marrakech, Morocco \label{ouka}}

\author[0000-0003-1462-7739]{Micha\"{e}l Gillon} 
\affiliation{Astrobiology Research Unit, Universit\'e de Li\`ege, 19C All\'ee du 6 Ao\^ut, 4000 Li\`ege, Belgium}

\author{Robert~F.~Goeke} 
\affiliation{Department of Physics and Kavli Institute for Astrophysics and Space Research, Massachusetts Institute of Technology, Cambridge, MA 02139, USA}

\author[0000-0003-1728-0304]{Keith Horne}
\affiliation{SUPA Physics and Astronomy, University of St. Andrews, Fife, KY16 9SS Scotland, UK}

\author{Jonathan~M.~Irwin}
\affiliation{Institute of Astronomy, University of Cambridge, Madingley
  Road, Cambridge, CB3 0HA, United Kingdom}

\author[0000-0001-8923-488X]{Emmanuel Jehin} 
\affiliation{Space sciences, Technologies and Astrophysics Research (STAR) Institute, Universit\'e de Li\`ege, Belgium}

\author[0000-0002-4715-9460]{Jon~M.~Jenkins} 
\affiliation{NASA Ames Research Center, Moffett Field, CA 94035, USA}

\author[0000-0001-9911-7388]{David~W.~Latham} 
\affiliation{Center for Astrophysics \textbar\ Harvard \& Smithsonian, 60 Garden
  St., Cambridge, MA 02138, USA}

\author{Dan Moldovan} 
\affiliation{Google, Cambridge, MA, USA}

\author[0000-0001-9087-1245]{Felipe Murgas}
\affiliation{Univ. Grenoble Alpes, CNRS, IPAG, F-38000 Grenoble, France}
\affiliation{Instituto de Astrofísica de Canarias (IAC), E-38200 La Laguna, Tenerife, Spain}
\affiliation{Dept. Astrofísica, Universidad de La Laguna (ULL), E-38206 La Laguna, Tenerife, Spain}

\author[0000-0003-1572-7707]{Francisco~J.~Pozuelos} 
\affiliation{Astrobiology Research Unit, Universit\'e de Li\`ege, 19C All\'ee du 6 Ao\^ut, 4000 Li\`ege, Belgium}
\affiliation{Space sciences, Technologies and Astrophysics Research (STAR) Institute, Universit\'e de Li\`ege, Belgium}
\affiliation{Instituto de Astrof\'isica de Andaluc\'ia (IAA-CSIC), Glorieta de la Astronom\'ia s/n, 18008 Granada, Spain}

\author[0000-0003-2058-6662]{George~R.~Ricker} 
\affiliation{Department of Physics and Kavli Institute for Astrophysics and Space Research, Massachusetts Institute of Technology, Cambridge, MA 02139, USA}

\author[0000-0001-8227-1020]{Richard P. Schwarz}
\affiliation{Center for Astrophysics \textbar\ Harvard \& Smithsonian, 60 Garden
  St., Cambridge, MA 02138, USA}

\author[0000-0002-6892-6948]{S.~Seager} 
\affiliation{Department of Physics and Kavli Institute for Astrophysics and Space Research, Massachusetts Institute of Technology, Cambridge, MA 02139, USA}
\affiliation{Department of Earth, Atmospheric and Planetary Sciences, Massachusetts Institute of Technology, Cambridge, MA 02139, USA}
\affiliation{Department of Aeronautics and Astronautics, MIT, 77 Massachusetts Avenue, Cambridge, MA 02139, USA}

\author{Gregor Srdoc}
\affiliation{Kotizarovci Observatory, Sarsoni 90, 51216 Viskovo, Croatia}

\author{Stephanie Striegel}
\affiliation{ San Jose State University, 1 Washington Sq, San Jose, CA 95192, USA}

\author{Mathilde~Timmermans} 
\affiliation{Astrobiology Research Unit, Universit\'e de Li\`ege, 19C All\'ee du 6 Ao\^ut, 4000 Li\`ege, Belgium}

\author[0000-0001-7246-5438]{Andrew~Vanderburg} 
\affiliation{Department of Physics and Kavli Institute for Astrophysics and Space Research, Massachusetts Institute of Technology, Cambridge, MA 02139, USA}

\author[0000-0001-6763-6562]{Roland~Vanderspek} 
\affiliation{Department of Physics and Kavli Institute for Astrophysics and Space Research, Massachusetts Institute of Technology, Cambridge, MA 02139, USA}

\author[0000-0002-4265-047X]{Joshua~N.~Winn} 
\affiliation{Department of Astrophysical Sciences, Princeton University, 4 Ivy Lane, Princeton, NJ 08544, USA}
 
\begin{abstract}
We present the discovery of TOI-3235 b, a short-period Jupiter orbiting an M-dwarf with a stellar mass close to the critical mass at which stars transition from partially to fully convective. TOI-3235 b was first identified as a candidate from \textit{TESS} photometry, and confirmed with radial velocities from ESPRESSO, and ground-based photometry from HATSouth, MEarth-South, TRAPPIST-South, LCOGT, and ExTrA. We find that the planet has a mass of $\mathrm{\hatcurPPm\,\mjup}$ and a radius of $\mathrm{\hatcurPPr\,\rjup}$.  It orbits close to its host star, with an orbital period of $\mathrm{\hatcurLCPshort\,d}$, but has an equilibrium temperature of $\mathrm{\approx \hatcurPPteffshort \, K}$, well below the expected threshold for radius inflation of hot Jupiters. The host star has a mass of $\mathrm{\hatcurISOm{}\,\msun}$, a radius of $\mathrm{\hatcurISOrlong{}\,\rsun}$, an effective temperature of $\mathrm{\hatcurISOteffshort \, K}$, and a J-band magnitude of $\mathrm{\hatcurCCtwomassJmag}$. Current planet formation models do not predict the existence of gas giants such as TOI-3235 b around such low-mass stars. With a high transmission spectroscopy metric, TOI-3235 b is one of the best-suited giants orbiting M-dwarfs for atmospheric characterization.
\end{abstract}



\section{Introduction} \label{sec:intro}

While planets around M-dwarf stars are extremely abundant \citep[e.g.][]{dressing:2015, hirano:2018, mulders:2018, hsu:2020}, the vast majority of these planets are smaller than Neptune, particularly around less massive M-dwarfs ($M<0.5 \msun$). Standard core-accretion formation models have long predicted few Jovian-mass planets around these less massive M-dwarfs \citep[e.g.][who also anticipate a particular scarcity of short-period giant planets]{laughlin:2004}. More recent implementations such as the Bern model \citep{Burn:2021} reproduce the low-mass planet population very well, but predict few gas giants around all M-dwarfs, and cannot produce them around later M-dwarfs with $M<0.5 \msun$ without fine-tuning of the planetary migration \citep{schlecker:2022}. Even prior to the \textit{Transiting Exoplanet Survey Satellite} mission \citep[\textit{TESS},][]{ricker:2015}, there were discoveries that challenged this (such as Kepler-45 b, \citealt{johnson:2012}; HATS-6 b, \citealt{hartman:2015:hats6}; NGTS-1 b, \citealt{bayliss:2018}).  More recently, both \textit{TESS} and radial velocity (RV) surveys have added to the known giant planets orbiting low-mass stars (e.g. GJ 3512 b, \citealt{morales:2019}; TOI-3884 b, \citealt{almenara:2022}), suggesting a potential alternative formation pathway such as gravitational instability \citep[e.g.][]{boss:2006}. However, as noted by \cite{schlecker:2022}, gravitational instability is expected to form very massive planets of $\approx 10 \mjup$ on large orbits, while the planets found to date are mainly of Jupiter mass and many have short orbital periods. Likewise, most of these planets orbit early M-dwarfs, for which the Bern model can, though rarely, produce gas giants; the first, and until now only, exception was TOI-5205 b \citep{kanodia:2022}, which orbits an M4 star. It is also worth noting that the Bern models normally assume a smooth initial gas surface density distribution in the protoplanetary disk; a non-smooth density distribution could modify the migration history and potentially facilitate the formation of these planets.  

In this context, the discovery and characterization of giant planets around M-dwarfs, particularly later M-dwarfs, is of paramount importance to planetary formation and migration theory. Transiting planets confirmed by radial velocities, for which both the mass and radius can be measured, are especially valuable. In this letter, we present the transiting gas giant TOI-3235 b, orbiting an M4 star with a period of $\hatcurLCPshort$ days. It is only the second gas giant found to orbit a later M-dwarf on the boundary between partially and fully convective M-dwarfs \citep{chabrier:1997}, and is one of a mere dozen giant planets orbiting M-dwarf stars. The planet was first identified as a candidate by the \textit{TESS} mission, and confirmed with ground-based photometry from HATSouth, MEarth-South, TRAPPIST-South, LCOGT, and ExTrA, and RVs from ESPRESSO. 

We present the data in Sect. \ref{sec:obs}. The analysis is described in Sect. \ref{sec:analysis}. Finally, we discuss and summarize our findings in Sect. \ref{sec:disc}.

\section{Observations} \label{sec:obs}

\subsection{Photometry}
\subsubsection{TESS}

TOI-3235 was observed by the \textit{TESS} primary and extended missions, in sectors 11 (23rd April to 20th May 2019) and 38 (29th April to 26th May 2021) respectively. In both cases, it was observed with camera 2 and CCD 4. The long-cadence data (30-minute cadence for sector 11, 10-minute cadence for sector 38) were initially processed by the Quick-Look Pipeline \citep[QLP, ][]{huang:2020, huang:2020b}, which uses full-frame images (FFI) calibrated by the \texttt{tica} package \citep{fausnaugh:2020}. The QLP detected a planet and it was promoted to a TOI following \cite{guerrero:2012}, as noted in the ExoFOP archive \footnote{Located at \url{https://exofop.ipac.caltech.edu/tess/target.php?id=243641947}}. For our analysis, we downloaded the \textit{TESS} PDCSAP light curves \citep{stumpe:2012, smith:2012, stumpe:2014} processed by the \textit{TESS} Science Processing Operation Center pipeline \citep[SPOC,][]{jenkinsSPOC2016} at NASA Ames Research Center, from the TESS-SPOC High Level Science Product on MAST \citep{caldwell:2020}. The SPOC difference image centroiding analysis locates the source of the transit signal to within $3.3 \pm 2.5"$ of the target star \citep{Twicken:DVdiagnostics2018}. The \textit{TESS} light curves are shown in Figure \ref{fig:tess}, and the data listed in Table \ref{tab:phfu}. 

\subsubsection{HATSouth}
HATSouth \citep{bakos:2013:hatsouth} is a network of 24 telescopes, distributed in three sites at Las Campanas Observatory (LCO) in Chile, the site of the H.E.S.S. gamma-ray observatory in Namibia, and Siding Spring Observatory (SSO) in Australia. Each telescope has a \tsize{0.18} aperture and \ccdsize{4K} front-illuminated CCD cameras. HATSouth observed TOI-3235 from 11th February 2017 through 15th May 2017, from all three sites. The data were reduced as described in \cite{penev:2013:hats1}. The transit was clearly detected, but was not flagged by the automated search due to the high transit depth and the pre-Gaia poor constraint on the stellar size from J-K magnitudes. The light curve is shown in Figure \ref{fig:toi3235-ground-phot} (left panel), and the data are listed in Table \ref{tab:phfu}. 

\subsubsection{MEarth-South}
MEarth-South is an array of eight \tsize{0.4} telescopes at the Cerro Tololo Inter-American Observatory (CTIO) in Chile \citep{nutzman:2008, irwin:2015}. M-Earth observed TOI-3235 with six telescopes on 21st June 2021 in the RG715 filter with $\rm{60\, s}$ exposure time, obtaining a full transit of TOI-3235.01. The light curves are shown in Figure \ref{fig:toi3235-ground-phot} (right panel), where the data from all six telescopes have been plotted together, and the data are listed in Table \ref{tab:phfu}.

\subsubsection{TRAPPIST-South}
TRAPPIST-South \citep{jehin:2011,Gillon2011} is a \tsize{0.6} Ritchey-Chretien robotic telescope at La Silla Observatory in Chile, equipped with a \ccdsize{2K} back-illuminated CCD camera with a pixel scale of 0.65\arcsec/pixel, resulting a field of view of $22\arcmin\times22\arcmin$. A full transit of TOI-3235.01 was observed by TRAPPIST-South on 10th May 2022 in the Sloan-$z'$ filter with an exposure time of 100s. We used the {\tt TESS Transit Finder} tool, which is a  customised version of the {\tt Tapir} software package \citep{jensen2013}, to schedule the observations. Data reduction and photometric measurement were performed using the {\tt PROSE}\footnote{\textit{PROSE}: \url{https://github.com/lgrcia/prose}} pipeline \citep{garcia2021}. The light curve is shown in Figure \ref{fig:toi3235-ground-phot} (right panel), and the data are listed in Table \ref{tab:phfu}.

\subsubsection{LCOGT}
The Las Cumbres Observatory global telescope network \citep[LCOGT,][]{brown:2013:lcogt} is a globally distributed network of \tsize{1} telescopes. The telescopes are equipped with $4096\times4096$ SINISTRO cameras having an image scale of $0\farcs389$ per pixel, resulting in a $26\arcmin\times26\arcmin$ field of view. TOI-3235 was observed by LCOGT with the SINISTRO instrument at the South Africa Astronomical Observatory (SAAO) site in the Sloan-$i'$ band on 10th June 2021, and at the Cerro Tololo Inter-American Observatory (CTIO) site in the Sloan-$g'$ band on 1st July 2022, full transits of TOI-3235.01 being obtained in both observations. We used the {\tt TESS Transit Finder}, which is a customized version of the {\tt Tapir} software package \citep{Jensen:2013}, to schedule our transit observations. The images were calibrated by the standard LCOGT {\tt BANZAI} pipeline \citep{McCully:2018}. The differential photometric data were extracted using {\tt AstroImageJ} \citep{Collins:2017}. The light curves are shown in Figure \ref{fig:toi3235-ground-phot} (right panel), and the data listed in Table \ref{tab:phfu}.

\subsubsection{ExTrA}
The ExTrA facility (Exoplanets in Transits and their Atmospheres, \citealt{bonfils:2015}) is composed of a near-infrared (0.85 to 1.55 $\mu$m) multi-object spectrograph fed by three \tsize{0.6} telescopes located at La Silla observatory in Chile. We observed 5 full transits of TOI-3235.01 on 2nd March 2022 (with three telescopes) and on 28th March 2022, 2nd April 2022, 23rd April 2022, and 24th May 2022 (with two telescopes). We observed using the fibers with $8\arcsec$ apertures, used the low resolution mode of the spectrograph (R$\sim20$) and 60-second exposures for all nights. At the focal plane of each telescope, five fiber positioners are used to pick the light from the target and four comparison stars. As comparison stars, we observed 2MASS J13493913-4615443, 2MASS J13515346-4623273, 2MASS J13510825-4612537 and 2MASS J13481046-4615434, with J-magnitude \citep{skrutskie:2006} and $T_{eff}$ \citep{gaiadr2} similar to TOI-3235. The resulting ExTrA data were analyzed using custom data reduction software. The light curves are shown in Figure \ref{fig:toi3235-ground-phot} (right panel), and the data listed in Table \ref{tab:phfu}.

\subsection{Radial Velocities}
\subsubsection{ESPRESSO}

ESPRESSO \citep[Echelle SPectrograph for Rocky Exoplanets and Stable Spectroscopic Observations,][]{pepe:2021} is an ultra-stable fibre-fed échelle high-resolution spectrograph installed at the incoherent combined Coudé facility of the Very Large Telescope (VLT) in Paranal Observatory, Chile. We observed TOI-3235 with ESPRESSO in HR mode (1 UT, $\mathrm{R \sim 140,000}$) between 2nd and 14th February 2022, obtaining 7 spectra under programme ID 108.22B4.001 aka 0108.C-0123(A). The spectra were reduced with the official ESPRESSO DRS v2.3.5 pipeline \citep{sosnowska:2015, modigliani:2020}, in the EsoReflex environment \citep{freudling:2013}. The RVs and bisector spans are listed in Table \ref{tab:rvs}, and the phase-folded RVs and bisector spans are shown in Figure \ref{fig:toi3235-RV-SED} (left panel). Two of the bisector spans are extreme outliers with values of $<-3000\, \ms$, and were excluded from the analysis.

\begin{deluxetable*}{lrrrrl}
\tabletypesize{\small}
\tablewidth{0pc}
\tablecaption{
    Light curve data for TOI-3235\label{tab:phfu}.
}
\tablehead{
    \colhead{BJD\tablenotemark{a}} & 
    \colhead{Mag\tablenotemark{b}} & 
    \colhead{\ensuremath{\sigma_{\rm Mag}}} &
    \colhead{Mag(orig)\tablenotemark{c}} & 
    \colhead{Filter} &
    \colhead{Instrument} \\
    \colhead{\hbox{~~~~(2,450,000$+$)~~~~}} & 
    \colhead{} & 
    \colhead{} &
    \colhead{} & 
    \colhead{} &
    \colhead{}
}
\startdata
$ 7853.13313 $ & $  13.60729 $ & $   0.03233 $ & $  -0.02073 $ & $ r$ & HATSouth/G701.3\\
$ 7879.05934 $ & $  13.62317 $ & $   0.02532 $ & $  -0.00485 $ & $ r$ & HATSouth/G701.3\\
$ 7801.28120 $ & $  13.57721 $ & $   0.03175 $ & $  -0.05081 $ & $ r$ & HATSouth/G701.3\\
$ 7853.13745 $ & $  13.66479 $ & $   0.03355 $ & $   0.03677 $ & $ r$ & HATSouth/G701.3\\
$ 7879.06378 $ & $  13.65081 $ & $   0.02521 $ & $   0.02279 $ & $ r$ & HATSouth/G701.3\\
$ 7801.28524 $ & $  13.61521 $ & $   0.03411 $ & $  -0.01281 $ & $ r$ & HATSouth/G701.3\\
$ 7801.29055 $ & $  13.67750 $ & $   0.05744 $ & $   0.04948 $ & $ r$ & HATSouth/G701.3\\
$ 7853.14366 $ & $  13.58338 $ & $   0.03234 $ & $  -0.04464 $ & $ r$ & HATSouth/G701.3\\
$ 7879.06997 $ & $  13.59425 $ & $   0.02449 $ & $  -0.03377 $ & $ r$ & HATSouth/G701.3\\
$ 7853.14794 $ & $  13.67704 $ & $   0.03710 $ & $   0.04902 $ & $ r$ & HATSouth/G701.3\\

\enddata
\tablenotetext{a}{
    Barycentric Julian Date computed on the TDB system with correction
	for leap seconds.
}
\tablenotetext{b}{
    The out-of-transit level has been subtracted.  For observations
    made with the HATSouth instruments these magnitudes have been corrected for
    trends using the EPD and TFA procedures applied {\em prior} to
    fitting the transit model.  This procedure may lead to an
    artificial dilution in the transit depths when used in its plain
    mode, instead of the signal reconstruction mode
    \citep{kovacs:2005:TFA}.  The blend factors for the HATSouth light
    curves are listed in Table~\ref{tab:planetparam}.  For observations
    made with follow-up instruments (anything other than ``HATSouth'' in the
    ``Instrument'' column), the magnitudes have been corrected for a
    quadratic trend in time, and for variations correlated with up to
    three PSF shape parameters, fit simultaneously with the transit.
}
\tablenotetext{c}{
    Raw magnitude values without correction for the quadratic trend in
    time, or for trends correlated with the seeing. 
}
\tablecomments{
    This table is available in a machine-readable form in the online
    journal.  A portion is shown here for guidance regarding its form
    and content.
}
\end{deluxetable*}

\tabletypesize{\scriptsize}
    \begin{deluxetable}{rrrrrr}
\tablewidth{0pc}
\tablecaption{
    Relative radial velocities and bisector spans from ESPRESSO for
    \hatcurhtr{}.  \label{tab:rvs}
}
\tablehead{
    \colhead{BJD} &
    \colhead{RV\tablenotemark{a}} &
    \colhead{\ensuremath{\sigma_{\rm RV}}\tablenotemark{b}} &
    \colhead{BS} &
    \colhead{\ensuremath{\sigma_{\rm BS}}} &
    \colhead{Phase}\\
    \colhead{\hbox{(2,450,000$+$)}} &
    \colhead{(\ms)} &
    \colhead{(\ms)} &
    \colhead{(\ms)} &
    \colhead{(\ms)} &
    \colhead{}
}
\startdata
 $ 9612.80279 $ & $  -185.54 $ & $     4.18 $ & \nodata      & \nodata      & $   0.224 $   \\
 $ 9613.81343 $ & $   126.16 $ & $     3.91 $ & \nodata      & \nodata      & $   0.613 $   \\
 $ 9615.71703 $ & $  -137.66 $ & $     5.29 $ & $   38.0 $ & $   10.6 $ & $   0.348 $ \\
 $ 9617.71593 $ & $  -121.86 $ & $     4.26 $ & $   32.6 $ & $    8.5 $ & $   0.119 $   \\
 $ 9618.70340 $ & $     0.95 $ & $     5.52 $ & $  -27.3 $ & $   11.1 $ & $   0.499 $   \\
 $ 9619.86092 $ & $    53.75 $ & $     3.50 $ & $  -12.7 $ & $    7.0 $ & $   0.946 $   \\
 $ 9624.86523 $ & $   129.76 $ & $     4.23 $ & $  -13.9 $ & $    8.4 $ & $   0.876 $   \\

\enddata
\tablenotetext{a}{
    The zero-point of these velocities is arbitrary. An overall offset
    $\gamma$ fitted to the velocities has been subtracted.
}
\tablenotetext{b}{
    Internal errors excluding the component of astrophysical jitter
    considered in \refsecl{analysis}.
}
    \end{deluxetable}

\section{Analysis} \label{sec:analysis}

\begin{figure*}[!ht]
{
 \centering
 \leavevmode
 \includegraphics[width={1.0\linewidth}]{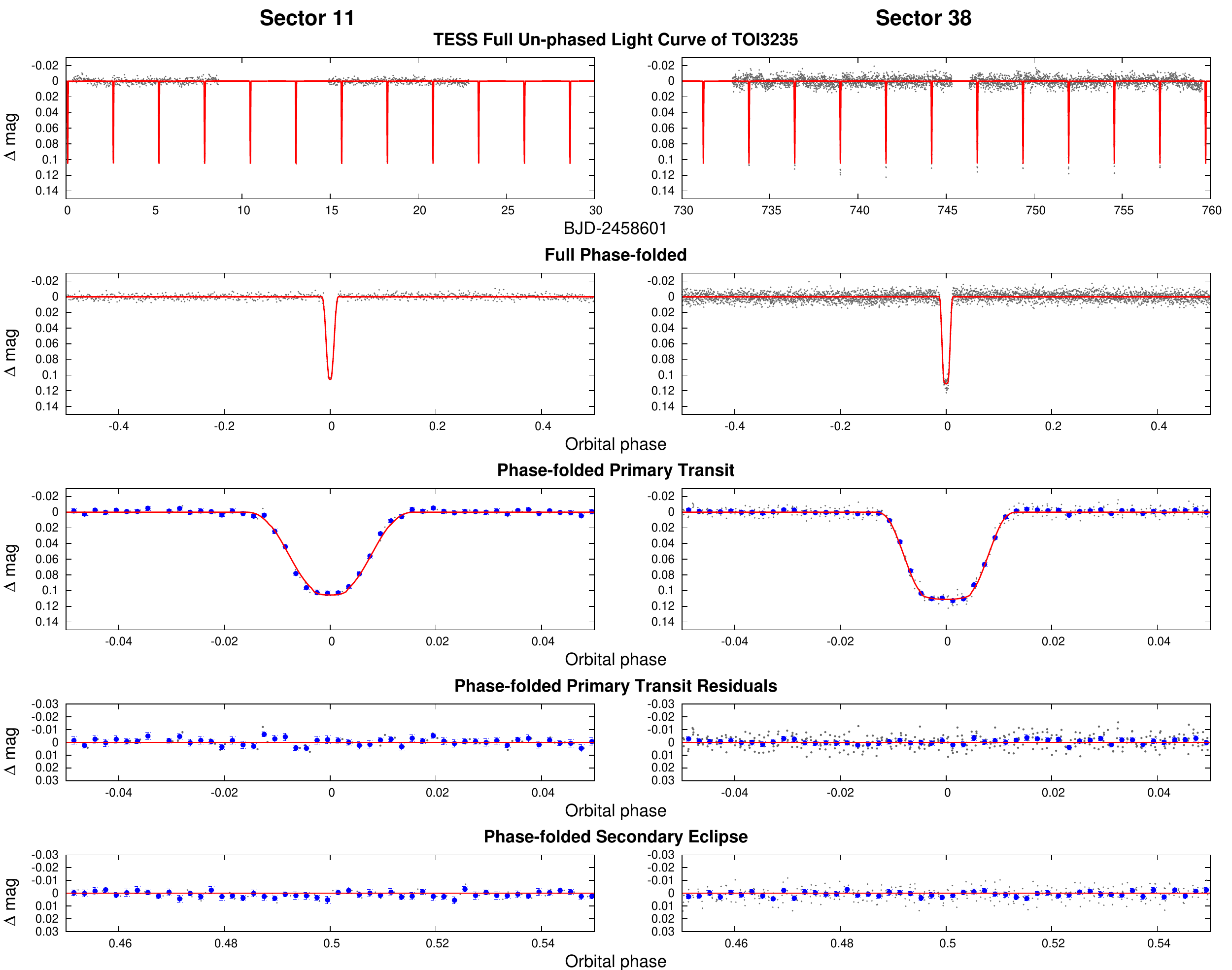}
}
\caption{
    {\em TESS} long-cadence light curves for TOI-3235, for sector 11 (left, 30-minute cadence) and sector 38 (right, 10-minute cadence). For each sector, we show the full un-phased light curve as a function of time ({\em top}), the full phase-folded light curve ({\em second}), the phase-folded light curve zoomed-in on the planetary transit ({\em third}), the residuals from the best-fit model, phase-folded and zoomed-in on the planetary transit ({\em fourth}), and the phase-folded light curve zoomed-in on the secondary eclipse ({\em bottom}). The solid red line in each panel shows the model fit to the light curve. The blue filled circles show the light curve binned in phase with a bin size of 0.002. Other observations included in our analysis of this system are shown in Figures~\ref{fig:toi3235-ground-phot} and ~\ref{fig:toi3235-RV-SED}. 
\label{fig:tess}
}
\end{figure*}

\begin{figure*}[!ht]
 {
 \centering
 \leavevmode
 \includegraphics[width={0.5\linewidth}]{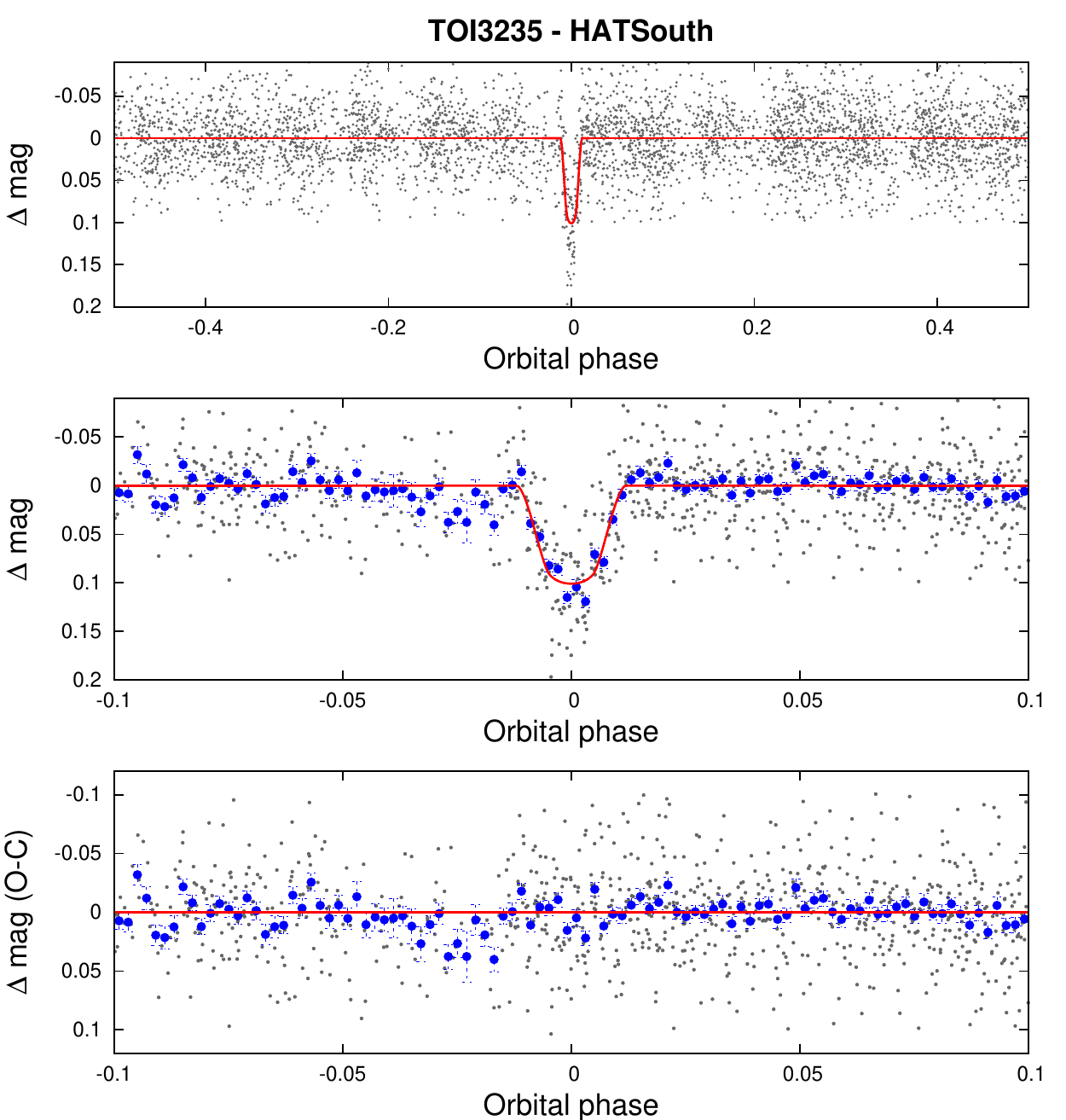}%
 \hfil
 \includegraphics[width={0.5\linewidth}]{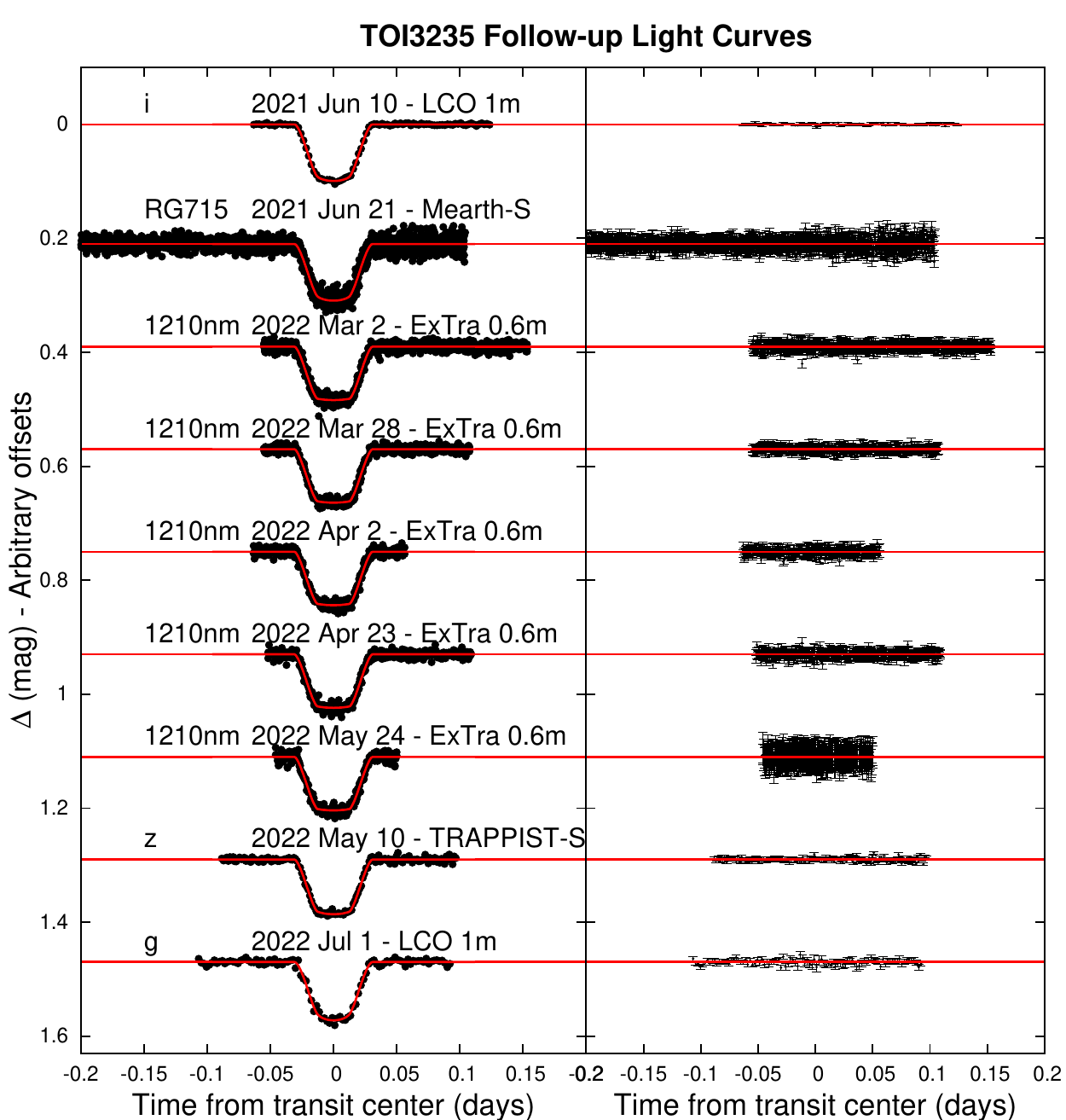}%
 }
                         
\caption{
    Ground-based photometry for the the transiting planet system TOI-3235. {\em Left:} Phase-folded unbinned full HATSouth light curve (top), light curve zoomed-in on the transit (middle), and residuals from the best-fit model zoomed-in on the transit (bottom). Solid red lines show the best-fit model. Blue circles show the light curves binned in phase with a bin size of 0.002. {\em Right:} Unbinned follow-up transit light curves corrected for instrumental trends fitted simultaneously with the transit model, which is overplotted (left), and residuals to the fit (right). Dates, filters and instruments are indicated. For ExTrA we indicate the midpoint of the spectral range. The error bars represent the photon
    and background shot noise, plus the readout noise. 
\label{fig:toi3235-ground-phot}
}
\end{figure*}

\begin{figure*}[!ht]
 {
 \centering
 \leavevmode
 \includegraphics[width={0.5\linewidth}]{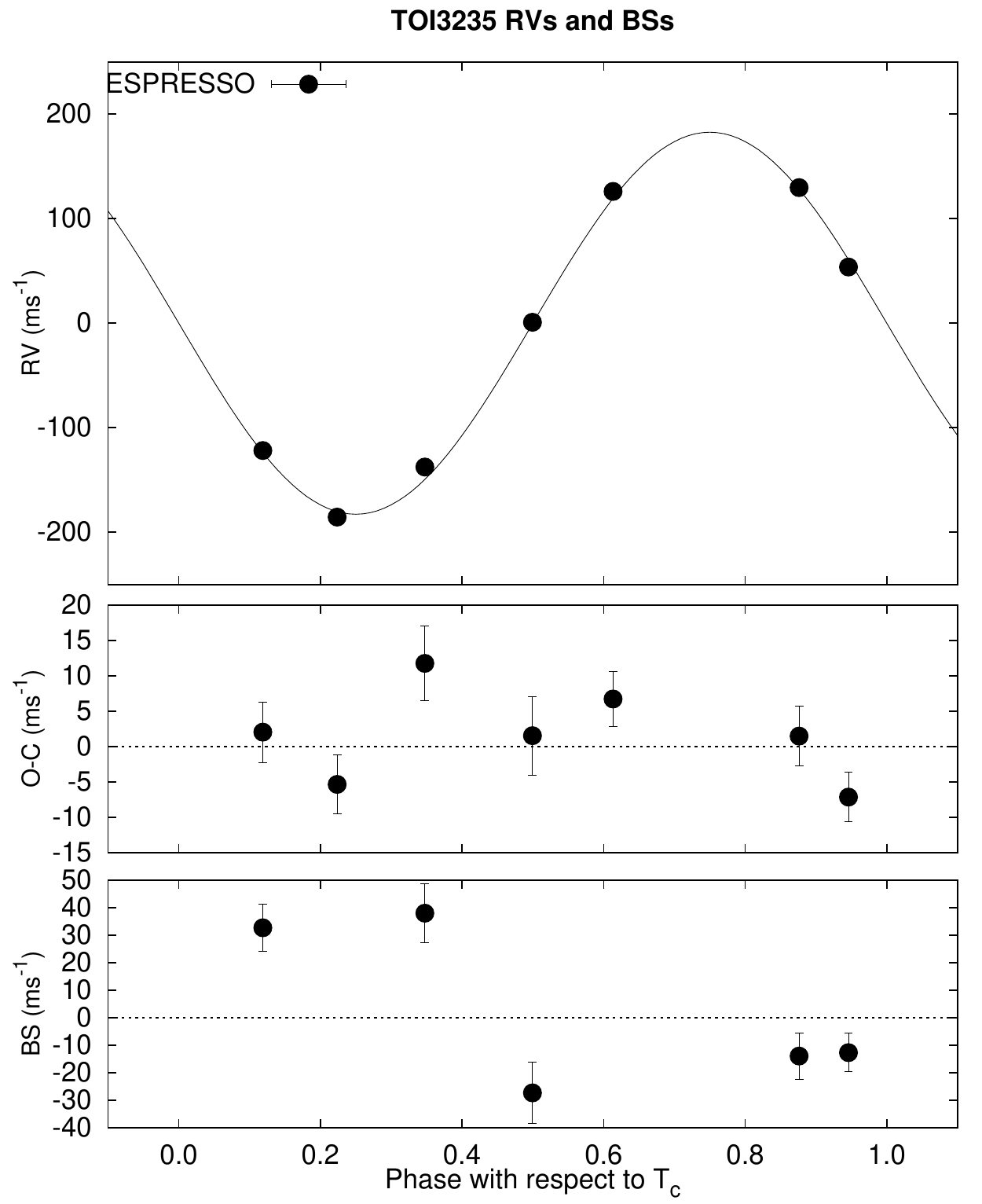}%
 \hfil
 \includegraphics[width={0.5\linewidth}]{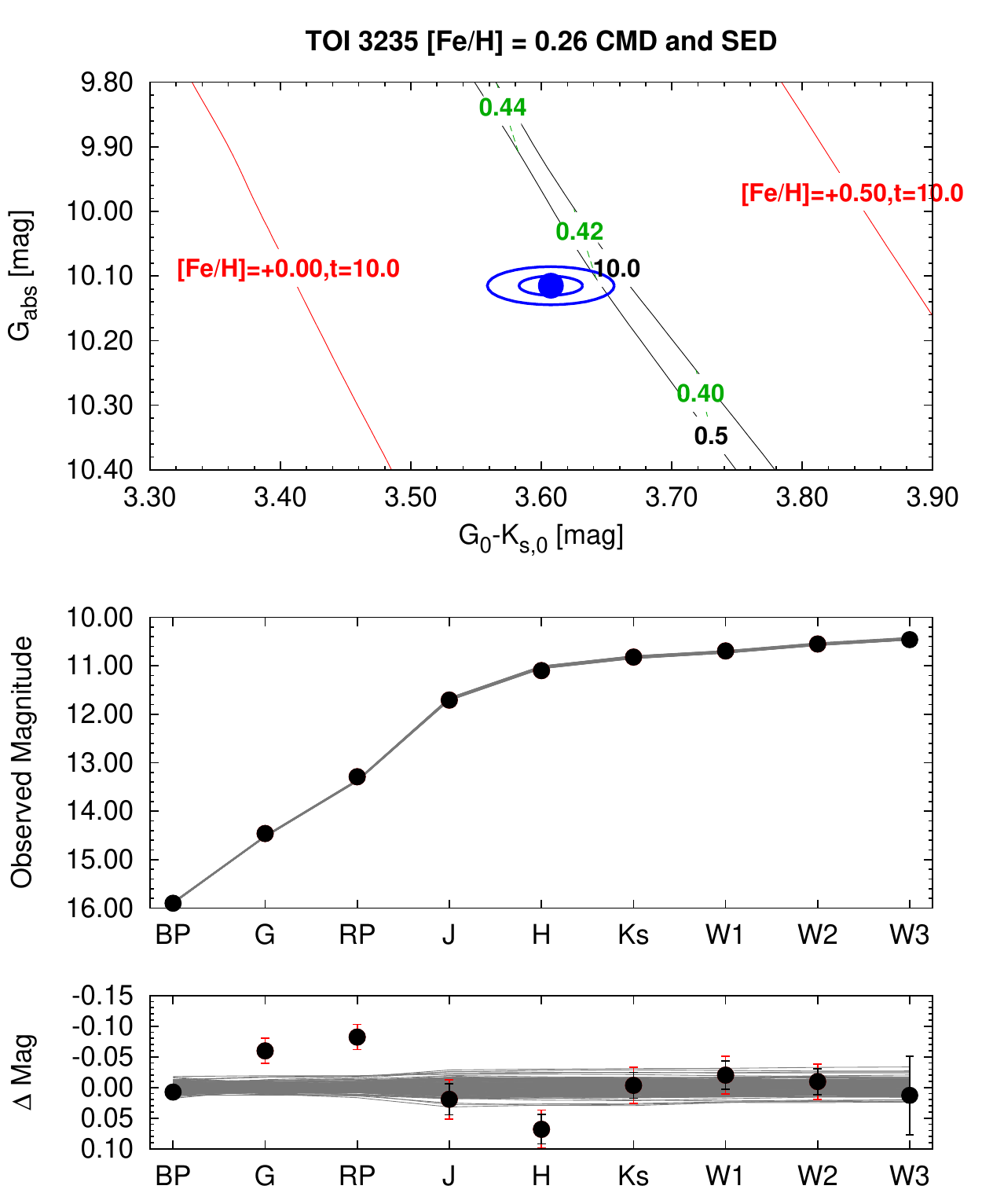}%
 }                        
\caption{
{\em Left:}
High-precision RVs from ESPRESSO/VLT phased with respect to the mid-transit time, together with the best-fit model, where the center-of-mass velocity has been subtracted (top); RV $O\!-\!C$ residuals (centre); and bisector spans (bottom). Error bars include the estimated jitter, which is a free parameter in the fitting.
{\em Top Right:} Absolute $G$ magnitude vs.\ the de-reddened $G - K_{S}$ color from Gaia DR2 and 2MASS (filled blue circle) and $1\sigma$ and $2\sigma$ confidence regions, including estimated systematic errors in the photometry (blue lines), compared to theoretical isochrones (black lines, ages listed in Gyr) and stellar evolution tracks (green dashed lines, mass listed in solar masses) from the MIST models interpolated at the best-estimate value for the host metallicity. The red lines show isochrones at higher and lower metallicities than the best-estimate value, labelled with their metallicity and age in Gyr. {\em Bottom Right:} SED as measured via broadband photometry through the listed filters (top), and $O\!-\!C$ residuals from the best-fit model (bottom). We plot the observed magnitudes without correcting for distance or extinction. Overplotted are 200 model SEDs randomly selected from the MCMC posterior distribution produced through the global analysis (gray lines). Black error bars show the catalog errors for the broad-band photometry measurements; red error bars add an assumed 0.02\,mag systematic uncertainty in quadrature to the catalog errors. These latter uncertainties are used in the fit.
\label{fig:toi3235-RV-SED}}
\end{figure*}

We carried out a joint analysis of the photometric, astrometric and RV data for TOI-3235\,b following the methods of \citet{hartman:2019:hats6069} and \citet{bakos:2020:hats71}. We fit the light curve data shown in Figures~\ref{fig:tess} and~\ref{fig:toi3235-ground-phot}, together with the broad-band catalog photometry and {\em Gaia} parallax measurement listed in Table~\ref{tab:stellarobserved}, and the RV data shown in Figure~\ref{fig:toi3235-RV-SED}. The model also makes use of the predicted absolute magnitudes in each bandpass from the MIST isochrones and of the extinction, constrained from the SED. We use a \citet{mandel:2002} transit model with quadratic limb darkening to fit the light curves and assume a Keplerian orbit for fitting the RV measurements. The limb darkening coefficients are allowed to vary, with priors based on the \citet{claret:2012,claret:2013,claret:2018} theoretical models. The stellar parameters are constrained using isochrones from version 1.2 of the MIST theoretical stellar evolution models \citep{paxton:2011,paxton:2013,paxton:2015,choi:2016,dotter:2016}.  We allow the line of sight extinction $A_{V}$ to vary in the fit, imposing a maximum of $0.527$\,mag and a Gaussian prior of $0.055 \pm 0.2$\,mag based on the MWDUST 3D Galactic extinction model \citep{bovy:2016}.

We used the ODUSSEAS software \citep{antoniadis:2020}, developed specifically for M-dwarfs, to measure the $\feh$ and $\teffstar$ from the ESPRESSO spectra. Although ODUSSEAS was developed for spectra with resolutions from $48\,000$ to $115\,000$, it has been successfully used with ESPRESSO spectra at their original $140\,000$ resolution \citep{lillo-box:2020}. We obtained preliminary values of $\feh = -0.0024 \pm 0.104$, $\teffstar = 3196 \pm 67$\,K, which were used as priors for the joint analysis\footnote{An independent estimate of $\teffstar=3421\pm53$\,K can be obtained using the absolute G magnitude $M_G$ from Equation~(11) of \citet{rabus:2019}, which is consistent at $\approx 2 \sigma$ with the value inferred from ODUSSEAS.}, in which a combination of the MIST evolution models, the transit-derived stellar bulk density, and the broad-band catalog photometry and parallax are employed to precisely constrain the host star parameters. To determine the spectral type, we used the PyHammer tool \citep{roulston:2020} with the ESPRESSO spectra, which returned an M5 spectral type. However, colour index comparisons with the tables of \cite{pecaut:2013} suggest an earlier spectral type of M3-M4, and visual inspection with the `eyecheck' facility of PyHammer shows an M4 template is also a good match to the spectrum. Therefore, we adopt an M4 spectral type.

We modelled the observations both assuming a circular orbit for the planet, and allowing the orbit to have a non-zero eccentricity. We find that the free-eccentricity model produces an eccentricity consistent with zero ($e \hatcurRVeccentwosiglimeccen{}$ at 95\% confidence). A very low eccentricity is expected, given that we estimate a rapid tidal circularization timescale for this system of $\mathrm{\sim 6\, Myr}$ \citep{hut:1981}. We therefore adopt the parameters that result from assuming a circular orbit. Applying the transit least squares \citep[TLS, ][]{hippke:2019} algorithm to the HATSouth and {\em TESS} light curve residuals to the best-fit model finds no additional transit signals. The stellar parameters derived from the analysis assuming a circular orbit are listed in Table~\ref{tab:stellarderived}, while the planetary parameters are listed in Table~\ref{tab:planetparam}. The best-fit model is shown in Figs. \ref{fig:tess}, \ref{fig:toi3235-ground-phot}, and \ref{fig:toi3235-RV-SED}. We note that the light curve uncertainties are scaled up in the fitting procedure to achieve a reduced $\chi^2$ of unity, but the uncertainties shown in Fig. \ref{fig:toi3235-ground-phot} have not been scaled.

The resulting $\sim 1$\% and $\sim 0.5$\% respective uncertainties on the derived stellar mass and radius are well below the respective $\sim 5$\% and $\sim 4.2$\% estimated systematic uncertainties of \cite{tayar:2022} for these parameters, which stem from inaccuracies in the fundamental observables and stellar evolution models. Likewise, the formal uncertainties of $7.4$\,K on the posterior stellar effective temperature and 0.017\,dex on the metallicity are likely quite a bit smaller than the systematic uncertainties, which we may expect to be closer to the ODUSSEAS-derived uncertainties of $\sim 70$\,K and $\sim 0.1$\,dex, respectively. 
However, as described in \citep{eastman:2022}, uncertainties smaller than the general error floors of Tayar et al. (2022) can be achieved for transiting planets by measuring the stellar density $\rhostar$ directly from the transit and employing it in the derivation of other stellar parameters.
Although our fit self-consistently accounts for the relation between the stellar density, transit parameters, $\mstar$, $\lstar$, $\rstar$, and $\teff$ throughout the fit, as suggested by \cite{eastman:2022}, and imposes a constraint that each link in the chain must match a stellar evolution model, it does not account for systematic errors in those models when imposing this constraint, and thus the formal uncertainties derived in this analysis are too small. Therefore, we conservatively adopt the error floors of \cite{tayar:2022}, which we report in brackets in Table \ref{tab:stellarderived}; for [Fe/H] we report the ODUSSEAS-derived uncertainty. These systematic uncertainties were formally propagated out to the planetary parameters. Regarding the planetary equilibrium temperature $T_{\rm eq}$ in particular, it is calculated under the assumptions of 0 albedo and full and instantaneous redistribution of heat, which are unlikely to hold completely in reality but provide a useful approximation.

The formal fit gives a young age of \hatcurISOage Gyr for the host star. However, this is primarily driven by the photometry being somewhat blue compared to the model values (see Fig. \ref{fig:toi3235-RV-SED}, top right), which are known to be uncertain for M-dwarfs. We see no other evidence of youth such as flares. Likewise, the GLS periodogram of the HATSouth photometry shows a significant peak at $44.4264 \pm 0.0010$ days; taking this as the stellar rotation period, the relations of \cite{engle:2018} suggest a much larger age of $\approx 2.7$ Gyr. We also used the BANYAN $\Sigma$ tool \citep{gagne:2018} to check the probability of TOI-3235 belonging to known young stellar associations given its Gaia DR3 \citep{gaiadr3} proper motions and radial velocity, finding it has a 99.9\% probability of being a field star.

Independent estimates of the stellar mass and radius can be obtained from the $\mathrm{K_S}$ magnitude using the mass-radius-luminosity relations of \cite{mann:2018} and \cite{rabus:2019}. Applying these relations leads to a mass of $\mstar = 0.3605 \pm 0.087 \msun$ and a radius of $\rstar = 0.37 \pm 0.07 \rsun$. While the radius is fully consistent with that obtained via global modelling, the mass is lower at $1.5\sigma$. We choose to adopt the values from the global modelling, since it accounts for all variables simultaneously. We also note that the planetary mass and radius calculated by employing the values obtained through the mass-radius-luminosity relations remain consistent with those computed from the global modelling values; thus, adopting the lower stellar mass from the mass-radius-luminosity relations would only make this giant planet even more unusual.

\begin{deluxetable*}{lcl}

\tablewidth{0pc}
\tabletypesize{\tiny}
\tablecaption{
    Astrometric, Spectroscopic and Photometric parameters for \hatcurhtr{}
    \label{tab:stellarobserved}
}
\tablehead{
    \multicolumn{1}{c}{~~~~~~~~Parameter~~~~~~~~} &
    \multicolumn{1}{c}{Value}                     &
    \multicolumn{1}{c}{Source}
}
\startdata
\noalign{\vskip -3pt}
\sidehead{Astrometric properties and cross-identifications}
~~~~2MASS-ID\dotfill               & \hatcurCCtwomass{}  & \\
~~~~TIC-ID\dotfill                 & \hatcurTICID{}  & \\
~~~~Gaia~DR3-ID\dotfill                 & \hatcurCCgaiadrthree{} & \\
~~~~R.A. (J2000)\dotfill            & \hatcurCCra{} & Gaia DR3\\
~~~~Dec. (J2000)\dotfill            & \hatcurCCdec{} & Gaia DR3\\
~~~~$\mu_{\rm R.A.}$ (\masy)              & \hatcurCCpmra{} & Gaia DR3\\
~~~~$\mu_{\rm Dec.}$ (\masy)              & \hatcurCCpmdec{} & Gaia DR3\\
~~~~parallax (mas)              & \hatcurCCparallax{}    & Gaia DR3\\
~~~~radial velocity (\kms)              & $-14.96 \pm 2.72$    & Gaia DR3\\
\sidehead{Spectroscopic properties}
~~~~$\teffstar$ (K)\dotfill         &  \hatcurSMEiteff{}   & ODUSSEAS/ESPRESSO\tablenotemark{a} \\
~~~~$\feh$\dotfill                  &  \hatcurSMEizfeh{}   &   ODUSSEAS/ESPRESSO\tablenotemark{a} \\
~~~~Spectral type \dotfill & M4 & this work\\
\sidehead{Photometric properties\tablenotemark{b}}
~~~~$G$ (mag)\dotfill               &  \hatcurCCgaiamGthree{}  & Gaia DR3 \\
~~~~$BP$ (mag)\dotfill               &  \hatcurCCgaiamBPthree{}  & Gaia DR3 \\
~~~~$RP$ (mag)\dotfill               &  \hatcurCCgaiamRPthree{}  & Gaia DR3 \\
~~~~$J$ (mag)\dotfill               &  \hatcurCCtwomassJmag{} & 2MASS           \\
~~~~$H$ (mag)\dotfill               &  \hatcurCCtwomassHmag{} & 2MASS           \\
~~~~$K_s$ (mag)\dotfill             &  \hatcurCCtwomassKmag{} & 2MASS           \\
~~~~$W1$ (mag)\dotfill             &  \hatcurCCWonemag{} & WISE           \\
~~~~$W2$ (mag)\dotfill             &  \hatcurCCWtwomag{} & WISE           \\
~~~~$W3$ (mag)\dotfill             &  \hatcurCCWthreemag{} & WISE           \\
\enddata
\tablenotetext{a}{
The ODUSSEAS-derived $\teffstar$ and $\feh$ are not the final adopted parameters, but are used as priors for the global modelling.
}
\tablenotetext{b}{
    The listed uncertainties for the photometry are taken from the catalogs. For the analysis we assume a systematic uncertainty floor of 0.02\,mag.
}
\end{deluxetable*}

\begin{deluxetable}{lc}
\tablewidth{0pc}
\tabletypesize{\footnotesize}
\tablecaption{
    Derived stellar parameters for \hatcurhtr{}
    \label{tab:stellarderived}
}
\tablehead{
    \multicolumn{1}{c}{~~~~~~~~Parameter~~~~~~~~} &
    \multicolumn{1}{c}{Value} 
}
\startdata
~~~~$\mstar$ ($\msun$)\dotfill      &  \hatcurISOmlong{} ($\pm 0.020$)    \\
~~~~$\rstar$ ($\rsun$)\dotfill      &  \hatcurISOrlong{} ($\pm0.016$)   \\
~~~~$\loggstar$ (cgs)\dotfill       &  \hatcurISOlogg{} ($\pm 0.063$)   \\
~~~~$\rhostar$ (\gcmc)\dotfill       &  \hatcurLCrho{}   \\
~~~~$\lstar$ ($\lsun$)\dotfill      &  \hatcurISOlum{} ($\pm 0.00039$)    \\
~~~~$\teffstar$ (K)\dotfill      &  \hatcurISOteff{} ($\pm 68$) \\
~~~~\feh\ (dex)\dotfill      &  \hatcurISOzfeh{} ($\pm 0.1$) \\
~~~~$A_{V}$ (mag)\dotfill               &  \hatcurXAv{}    \\
~~~~Distance (pc)\dotfill           &  \hatcurXdistred{}\phn  \\
\enddata
\tablecomments{
The listed parameters are those determined through the joint analysis described in Section~\ref{sec:analysis} assuming a circular orbit for the planet. The first uncertainties listed for each parameter are the statistical uncertainties from the fit, not including systematic errors. Values in brackets report the estimated uncertainty floors due to inaccuracies in the fundamental observables and/or the MIST stellar evolution models, where appropriate. These latter floors were formally propagated to the planetary parameter uncertainties.
}
\end{deluxetable}

\begin{deluxetable}{lc}
\tabletypesize{\tiny}
\tablecaption{Adopted orbital and planetary parameters for \hatcurhtr{}\,b\label{tab:planetparam}}
\tablehead{
    \multicolumn{1}{c}{~~~~~~~~~~~~~~~Parameter~~~~~~~~~~~~~~~} &
    \multicolumn{1}{c}{Value} 
}
\startdata
\noalign{\vskip -3pt}
\sidehead{\Lc{} parameters}
~~~$P$ (days)             \dotfill    & \hatcurLCP{} \\
~~~$T_c$ (${\rm BJD\_{}TDB}$)    
      \tablenotemark{a}   \dotfill    & \hatcurLCT{}  \\
~~~$T_{14}$ (days)
      \tablenotemark{a}   \dotfill    & $\hatcurLCdur{}$  \\
~~~$T_{12} = T_{34}$ (days)
      \tablenotemark{a}   \dotfill    & $\hatcurLCingdur{}$  \\
~~~$\arstar$              \dotfill    & $\hatcurPPar{}$  \\
~~~$\zrstar$ \tablenotemark{b}             \dotfill    & $\hatcurLCzeta$\phn \\
~~~$\rpl/\rstar$          \dotfill    & $\hatcurLCrprstar{}$ \\
~~~$b^2$                  \dotfill    & $\hatcurLCbsq{}$ \\
~~~$b \equiv a \cos i/\rstar$
                          \dotfill    & $\hatcurLCimp{}$\\
~~~$i$ (deg)              \dotfill    & $\hatcurPPi{}$\phn \\
\sidehead{Dilution factors \tablenotemark{c}}
~~~HAT G701/3 \dotfill & $\hatcurLCiblendA{}$ \\
~~~HAT G701/4 \dotfill & $\hatcurLCiblendB{}$ \\
~~~{\em TESS} Sector 11 \dotfill & $\hatcurLCiblendC{}$ \\
~~~{\em TESS} Sector 38 \dotfill & $\hatcurLCiblendD{}$ \\
\sidehead{Limb-darkening coefficients \tablenotemark{d}}
~~~$c_1,T$                  \dotfill    & $\hatcurLBiT{}$ \\
~~~$c_2,T$                  \dotfill    & $\hatcurLBiiT{}$ \\
~~~$c_1,g$                  \dotfill    & $\hatcurLBig{}$ \\
~~~$c_2,g$                  \dotfill    & $\hatcurLBiig{}$ \\
~~~$c_1,r$                  \dotfill    & $\hatcurLBir{}$ \\
~~~$c_2,r$                  \dotfill    & $\hatcurLBiir{}$ \\
~~~$c_1,z$                  \dotfill    & $\hatcurLBiz{}$ \\
~~~$c_2,z$                  \dotfill    & $\hatcurLBiiz{}$ \\
~~~$c_1,RG715$                  \dotfill    & $\hatcurLBii{}$ \\
~~~$c_2,RG715$                  \dotfill    & $\hatcurLBiii{}$ \\
\sidehead{RV parameters}
~~~$K$ (\ms)              \dotfill    & $\hatcurRVK{}$\phn\phn \\
~~~$e$ \tablenotemark{e}               \dotfill    & $\hatcurRVeccentwosiglimeccen$\\
~~~RV jitter ESPRESSO (\ms)        \dotfill    & $\hatcurRVjittertwosiglim{}$\\
\sidehead{Planetary parameters}
~~~$\mpl$ ($\mjup$)       \dotfill    & $\hatcurPPmlong{}$ \\
~~~$\rpl$ ($\rjup$)       \dotfill    & $\hatcurPPrlong{}$ \\
~~~$C(\mpl,\rpl)$
    \tablenotemark{g}     \dotfill    & $\hatcurPPmrcorr{}$ \\
~~~$\rhopl$ (\gcmc)       \dotfill    & $\hatcurPPrho{}$ \\
~~~$\log g_p$ (cgs)       \dotfill    & $\hatcurPPlogg{}$ \\
~~~$a$ (AU)               \dotfill    & $\hatcurPParel{}$ \\
~~~$T_{\rm eq}$ (K)         \dotfill   & $\hatcurPPteff{}$ \\
~~~$\Theta$ \tablenotemark{h} \dotfill & $\hatcurPPtheta{}$ \\
~~~$\log_{10}\langle F \rangle$ (cgs) \tablenotemark{i}
                          \dotfill    & $\hatcurPPfluxavglog{}$ \\
\enddata
\tablecomments{
We adopt a model in which the orbit is assumed to be circular. See the discussion in Section~\ref{sec:analysis}.
}
\tablenotetext{a}{
    Times are in Barycentric Julian Date calculated on the Barycentric Dynamical Time (TDB) system.
    \ensuremath{T_c}: Reference epoch of
    mid transit that minimizes the correlation with the orbital
    period.
    \ensuremath{T_{14}}: total transit duration, time
    between first to last contact;
    \ensuremath{T_{12}=T_{34}}: ingress/egress time, time between first
    and second, or third and fourth contact.
}
\tablenotetext{b}{
   Reciprocal of the half duration of the transit used as a jump parameter in our MCMC analysis in place of $\arstar$. It is related to $\arstar$ by the expression $\zrstar = \arstar(2\pi(1+e\sin\omega))/(P\sqrt{1-b^2}\sqrt{1-e^2})$ \citep{bakos:2010:hat11}.
}
\tablenotetext{c}{
    Scaling factor applied to the model transit fit to the HATSouth and {\em TESS} light curves. It accounts for dilution of the transit due to blending from neighboring stars, over-filtering of the light curve, or over-correction of dilution in the {\em TESS} SPOC light curves. These factors are varied in the fit, with independent values adopted for each light curve.
}
\tablenotetext{d}{
    Values for a quadratic law. The limb-darkening parameters were directly varied in the fit, using the tabulations from \cite{claret:2012,claret:2013,claret:2018} to place Gaussian priors on their values, assuming a prior uncertainty of $0.2$ for each coefficient.
}
\tablenotetext{e}{
    95\% confidence upper limit on the eccentricity determined when $\sqrt{e}\cos\omega$ and $\sqrt{e}\sin\omega$ are allowed to vary in the fit.
}
\tablenotetext{f}{
    Term added in quadrature to the formal RV uncertainties for each
    instrument. It is a free parameter in the fitting routine. 
}
\tablenotetext{g}{
    Correlation coefficient between the planetary mass \mpl\ and radius
    \rpl\ estimated from the posterior parameter distribution.
}

\tablenotetext{h}{
    The Safronov number is given by $\Theta = \frac{1}{2}(V_{\rm
    esc}/V_{\rm orb})^2 = (a/\rpl)(\mpl / \mstar )$
    \citep[see][]{hansen:2007}.
}
\tablenotetext{i}{
    Incoming flux per unit surface area, averaged over the orbit.
}
\end{deluxetable}

\section{Discussion and Conclusions}\label{sec:disc}

\begin{figure}[htb]
    \centering
    \includegraphics[width=1\hsize]{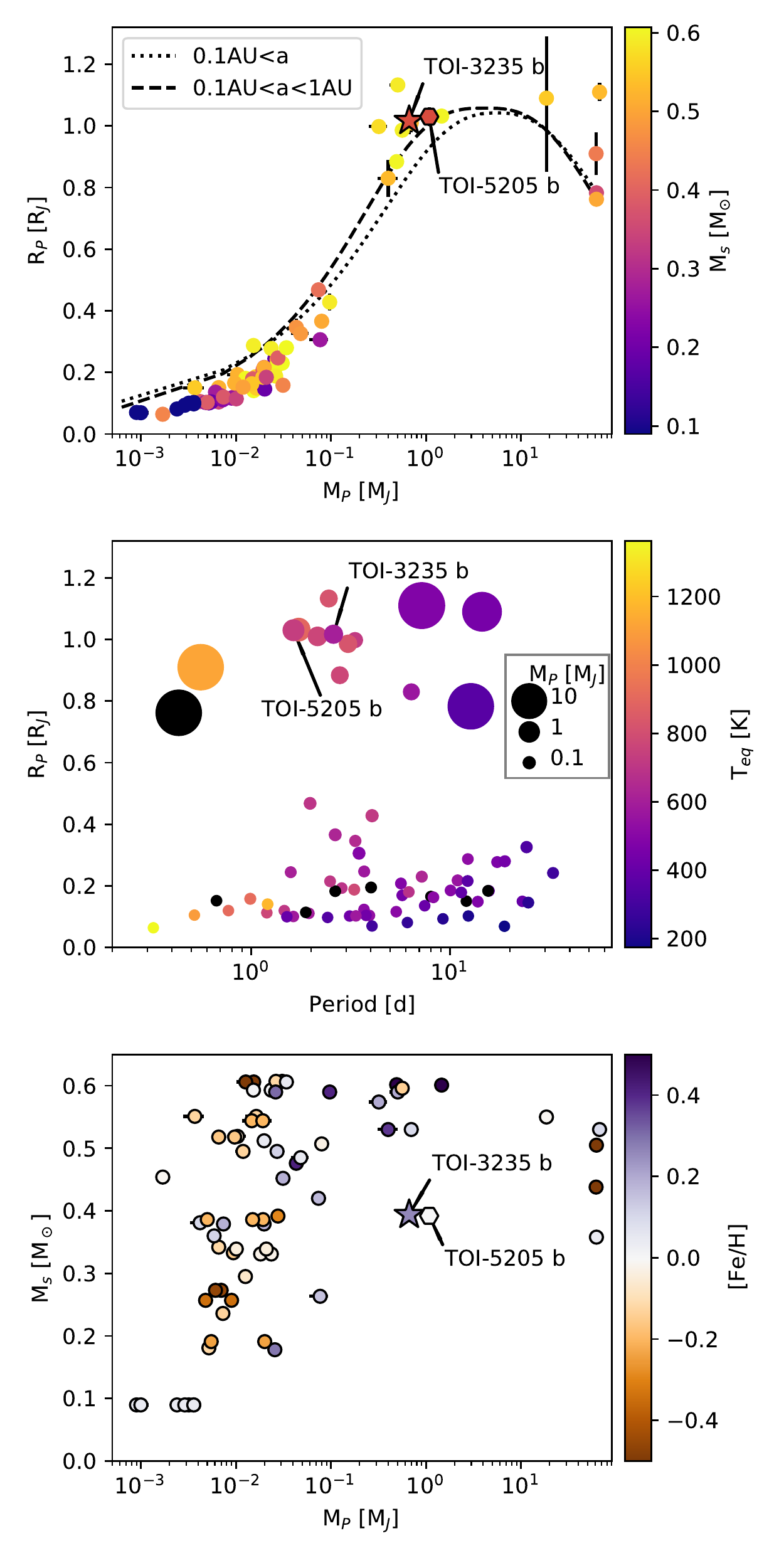}
    \caption{{\em Top}: Mass-radius diagram for M-dwarf planets with masses and radii measured to better than 25\%, as reported in TEPCAT. The markers are colour-coded by host star mass. TOI-3235 b and its analogue TOI-5205 b are plotted with star and hexagon symbols respectively and labelled. Theoretical mass-radius curves from \cite{mordasini:2012} are plotted with dashed and dotted lines. {\em Centre}: Period-radius diagram for the same planets. The markers are scaled by planet mass and colour-coded by equilibrium temperature (black when it could not be computed). TOI-3235 b and TOI-5205 b are labelled. {\em Bottom}: Planet mass vs. stellar mass diagram for the same planets. The markers are colour-coded by host star metallicity. TOI-3235 b and TOI-5205 b are plotted with star and hexagon symbols respectively and labelled.}
    \label{fig:rad-mass-period}
\end{figure}

TOI-3235 b is a close-in Jupiter with $\mpl = \hatcurPPm \, \mjup$, $\rpl = \hatcurPPr \, \rjup$, orbiting a $\hatcurISOm \, \msun$ M-dwarf with a period of $\hatcurLCP$ days. To place it in the context of the M-dwarf planet population, in Figure \ref{fig:rad-mass-period} we plot TOI-3235 b together with all other well-characterized planets from the TEPCAT catalogue \citep{southworth:2011} hosted by stars with $\mstar \leq 0.61\, \msun$ (limit chosen to include $\mstar \approx 0.6\, \msun$ stars on K-M boundary). In mass-radius space (Fig. \ref{fig:rad-mass-period}, top panel), TOI-3235 b joins a small cluster of ten giant planets with $0.8\, \rjup \leq \rpl \leq 1.2\, \rjup$, and $0.3\, \mjup \leq \mpl \leq 1.5\, \mjup$. Most of these planets (HATS-6 b, \citealt{hartman:2015:hats6}; HATS-74 b and HATS-75 b, \citealt{jordan:2022}; Kepler-45 b, \citealt{johnson:2012}; TOI-530 b, \citealt{gan:2022:toi530}; TOI-3714 b, \citealt{canas:2022}; WASP-80 b, \citealt{triaud:2013}) are hosted by early M-dwarfs with $\mstar \geq 0.5\,  \msun$. The sole other exception, aside from TOI-3235 b itself, is TOI-5205 b, a Jupiter-sized planet transiting a $0.392\, \msun$ star \citep{kanodia:2022}. Save for TOI-530 b, which has a somewhat longer period of $\mathrm{6.39\,d}$, these planets also cluster together in period-radius space (Fig. \ref{fig:rad-mass-period}, centre panel), forming a group of mid-range close-in Jupiters with periods between 1.63 and 3.33 days. Likewise, all their host stars except WASP-80 are metal-rich (Fig. \ref{fig:rad-mass-period}, bottom panel), in contrast to the wide range of metallicities shown by the host stars of the lower-mass planets. In particular, TOI-3235 has a metallicity of 0.26. These higher metallicities are consistent with previous findings \citep[e.g.][]{johnson:2009,rojas-ayala:2010} that M-dwarfs hosting giant planets tend to be metal-rich. The stellar mass vs planet mass diagram shown in this last panel also highlights the uniqueness of TOI-3235 b and its near twin TOI-5205 b, which inhabit an otherwise empty region of this parameter space.

Despite their clustering in mass-radius space, this group of giant planets spans a fairly wide range of densities, ranging from the very low-density HATS-6 b ($\mathrm{\rho \approx 0.4\, g/cm^3}$) to the Jupiter-analogue TOI-5205 b ($\mathrm{\rho \approx 1.3\, g/cm^3}$) and the high-density HATS-74 b ($\mathrm{\rho \approx 1.6\, g/cm^3}$). TOI-3235 b sits in the centre of the range, with $\mathrm{\rho \approx 0.78\, g/cm^3}$, comparable to the density of Saturn. They are all close to the peak of the theoretical mass-radius relationship derived by \cite{mordasini:2012} (Fig. \ref{fig:rad-mass-period}, top panel, where we show the relationships for both the full synthetic population, and for planets with $\mathrm{a < 1 au}$); save for TOI-530 b, HATS-74 b, and HATS-75 b, which sit on the curve for planets with $\mathrm{a < 1 au}$, and TOI-5205 b, which is consistent with it within error bars, all have larger radii than predicted. However, although all ten of these giants have periods shorter than the typical $\mathrm{10\, d}$ limit taken for hot Jupiters, they have equilibrium temperatures of $\mathrm{\approx 600-900\, K}$(Fig. \ref{fig:rad-mass-period}, centre panel), well below the $\mathrm{1000\, K}$ limit at which the incident flux is expected to begin to inflate the radii \citep{miller:2011, demory:2011, sarkis:2021}. It is also interesting to note that the low-mass planets generally have smaller radii than predicted, suggesting the theoretical relationship - derived from a synthetic population with a fixed stellar mass of $\mstar = 1\msun$ - may not be a good fit to M-dwarf planets overall. 

The similarities between these giant planets may point to similar formation and migration histories. However, the differences in host star mass indicate caution; we may be seeing two distinct populations, one corresponding to early M-dwarfs and one corresponding to later M-dwarfs. It is thus particularly interesting and relevant to compare TOI-3235 b to TOI-5205 b. Like TOI-5205, TOI-3235 sits on the edge of the Jao Gap, a narrow gap in the Hertzsprung–Russell Diagram first identified by \cite{jao:2018} in Gaia data and linked by the authors to the transition between partially and fully convective stars, with $\mathrm{M_G = 10.04 \pm 0.95}$ \citep{anders:2022}, and $\mathrm{B_P - R_P \approx 2.6}$ \citep{gaiadr3}. Both these stars are therefore in the transition region between partially and fully convective M-dwarfs, and as such are likely to undergo periodic changeovers from partially to fully convective and vice versa, that alter their radius and luminosity \citep[e.g.][]{vansaders:2012, baraffe:2018}. As noted by \cite{kanodia:2022}, these oscillations may impact the planetary orbital parameters and equilibrium temperature. It is possible that the similar planets of these similar stars may share similar formation and/or evolution histories. \cite{kanodia:2022} studied the disk mass necessary to form TOI-5205 b. Since the host stars have the same mass, we can extrapolate from their analysis; the main difference is that TOI-3235 b is rather less massive than TOI-5205 b. As regards planetary heavy-element mass, using the relations of \cite{thorngren:2016} we find a heavy-element mass of $\mathrm{M_Z \sim 45 M_\oplus}$, corresponding to 75\% of that of TOI-5205 b; therefore, assuming a solid core and scaling the results of \cite{kanodia:2022}, the required disk mass for TOI-3235 b becomes $\sim 2\%- 23\%$ the mass of the host star for $100\%-10\%$ formation efficiency respectively. While lower than the disk mass required to explain TOI-5205 b, given typical disk masses of the order of $\sim 0.1-5\%$ \citep{pascucci:2016}, the formation of TOI-3235 b still requires either an extremely high formation efficiency or a very massive disk.

TOI-3235 b also shows high potential for atmospheric characterization. We compute a Transmission Spectroscopy Metric (TSM, \citealt{kempton:2018}) of $\approx 160$, assuming a scale factor of 1.15. Comparing it to the group of M-dwarf planets it clusters with in mass-period-radius space, TOI-3235 b has the second-highest TSM, surpassed only by WASP-80 b (TSM $\approx 290$), and notably higher than its analogue TOI-5205 b (TSM $\approx 100$). Atmospheric characterization can help place constraints on the formation and migration history \citep[e.g.][]{hobbs:2022, molliere:2022} of this unexpected planet.

\vspace{5mm}
\facilities{\textit{TESS}, HATSouth, MEarth-South, TRAPPIST-South, LCOGT, ExTrA, ESPRESSO, Gaia, Exoplanet Archive}

\software{FITSH \citep{pal:2012}, BLS \citep{kovacs:2002:BLS}, VARTOOLS \citep{hartman:2016:vartools}, CERES \citep{brahm:2017:ceres}, ZASPE \citep{brahm:2017:zaspe}, ODUSSEAS \citep{antoniadis:2020}, AstroImageJ \citep{Collins:2017}, TAPIR \citep{Jensen:2013}}

\vspace{5mm}
We thank the referee for their helpful comments that improved this paper. 

This paper includes data collected by the \textit{TESS} mission, which are publicly available from the Mikulski Archive for Space Telescopes (MAST). The specific observations analyzed can be accessed via \dataset[10.17909/mdsd-2297]{https://doi.org/10.17909/mdsd-2297}. Funding for the \textit{TESS} mission is provided by NASA's Science Mission directorate. 

This research has made use of the Exoplanet Follow-up Observation Program website, which is operated by the California Institute of Technology, under contract with the National Aeronautics and Space Administration under the Exoplanet Exploration Program.

We acknowledge the use of public \textit{TESS} data from pipelines at the \textit{TESS} Science Office and at the \textit{TESS} Science Processing Operations Center.

Resources supporting this work were provided by the NASA High-End Computing (HEC) Program through the NASA Advanced Supercomputing (NAS) Division at Ames Research Center for the production of the SPOC data products. 

Based on observations collected at the European Organisation for Astronomical Research in the Southern Hemisphere under ESO programme  0108.C-0123(A).

A.J., R.B. and M.H. acknowledge support from ANID - Millennium Science Initiative - ICN12\_009. A.J. acknowledges additional support from FONDECYT project 1210718. R.B. acknowledges support from FONDECYT Project 1120075 and from project IC120009 “Millennium Institute of Astrophysics (MAS)” of the Millenium Science Initiative. This work was funded by the Data Observatory Foundation.

The MEarth Team gratefully acknowledges funding from the David and Lucile Packard Fellowship for Science and Engineering (awarded to D.C.). This material is based upon work supported by the National Science Foundation under grants AST-0807690, AST-1109468, AST-1004488 (Alan T. Waterman Award), and AST-1616624, and upon work supported by the National Aeronautics and Space Administration under Grant No. 80NSSC18K0476 issued through the XRP Program. This work is made possible by a grant from the John Templeton Foundation. The opinions expressed in this publication are those of the authors and do not necessarily reflect the views of the John Templeton Foundation.

The research leading to these results has received funding from the ARC grant for Concerted Research Actions, financed by the Wallonia-Brussels Federation. TRAPPIST is funded by the Belgian Fund for Scientific Research (Fond National de la Recherche Scientifique, FNRS) under the grant PDR T.0120.21. MG is F.R.S.-FNRS Research Director and EJ is F.R.S.-FNRS Senior Research Associate. Observations were carried out from ESO La Silla Observatory.

The postdoctoral fellowship of KB is funded by F.R.S.-FNRS grant T.0109.20 and by the Francqui Foundation.

This work makes use of observations from the LCOGT network. Part of the LCOGT telescope time was granted by NOIRLab through the Mid-Scale Innovations Program (MSIP). MSIP is funded by NSF.

Based on data collected under the ExTrA project at the ESO La Silla Paranal Observatory. ExTrA is a project of Institut de Plan\'etologie et d'Astrophysique de Grenoble (IPAG/CNRS/UGA), funded by the European Research Council under the ERC Grant Agreement n. 337591-ExTrA.
This work has been supported by a grant from Labex OSUG$@$2020 (Investissements d'avenir -- ANR10 LABX56).
This work has been carried out within the framework of the NCCR PlanetS supported by the Swiss National Science Foundation.
This work has been carried out within the framework of the National Centre of Competence in Research PlanetS supported by the Swiss National Science Foundation under grants 51NF40\textunderscore182901 and 51NF40\textunderscore205606. The authors acknowledge the financial support of the SNSF.

This publication benefits from the support of the French Community of Belgium in the context of the FRIA Doctoral Grant awarded to MT. 

The contributions at the Mullard Space Science Laboratory by E.M.B. have been supported by STFC through the consolidated grant ST/W001136/1.


\bibliography{hatsbib}{}
\bibliographystyle{aasjournal}



\end{document}